\pdfoutput=1
\UseRawInputEncoding
\documentclass[aps,prx,twocolumn,superscriptaddress,longbibliography]{revtex4-2}
\usepackage{graphicx}
\usepackage{amssymb,amsmath}
\usepackage{bm}
\usepackage{ bbold }
\usepackage{dcolumn}
\usepackage{float}
\usepackage[OT1]{fontenc} 
\usepackage{url}
\usepackage{mathrsfs}
\usepackage{slashed,comment}
\usepackage{color}
\usepackage{verbatim}
\usepackage{enumitem}
\usepackage{soul,physics}
\usepackage{float}    
\usepackage{amsmath}
\usepackage{booktabs} 
\usepackage{siunitx}  
\usepackage{multirow}
\usepackage[driverfallback=dvipdfm]{hyperref}
\hypersetup{pdfpagemode=FullScreen,colorlinks=true,breaklinks,urlcolor=blue,linkcolor=blue,citecolor=blue}

\usepackage{subfigure}
\usepackage{amssymb}
\usepackage{bm}
\usepackage{graphicx}
\usepackage{amsmath,amssymb}

\usepackage{pifont}

\begin{document}
 
  \title{Finite-Size Scaling of the Full Eigenstate Thermalization in Quantum Spin Chains}

  \author{Yuke Zhang}
  \affiliation{State Key Laboratory of Surface Physics \& Department of Physics, Fudan University, Shanghai, 200438, China}

  \author{Pengfei Zhang}
  \thanks{PengfeiZhang.physics@gmail.com}
  \affiliation{State Key Laboratory of Surface Physics \& Department of Physics, Fudan University, Shanghai, 200438, China}
  \affiliation{Hefei National Laboratory, Hefei 230088, China}

  \date{\today}

  \begin{abstract}
  Despite the unitary evolution of closed quantum systems, long-time expectation of local observables are well described by thermal ensembles, providing the foundation of quantum statistical mechanics. A promising route to understanding this quantum thermalization is the eigenstate thermalization hypothesis (ETH), which posits that individual energy eigenstates already appear locally thermal. Subsequent studies have extended this concept to the full ETH, which captures higher-order correlations among matrix elements through nontrivial relations. In this work, we perform a detailed exact-diagonalization study of finite-size corrections to these relations in the canonical ensemble. We distinguish two distinct sources of corrections: those arising from energy fluctuations, which decay polynomially with system size, and those originating from fluctuations within each energy window, which decay exponentially with system size. In particular, our analysis resolves the puzzle that, for certain observables, finite-size corrections exhibit anomalous growth with increasing system size even in chaotic systems. Our results provide a systematic and practical methodology for validating the full ETH in quantum many-body systems.
  \end{abstract}
    
  \maketitle

  \section{Introduction}
  Understanding how isolated quantum many-body systems approach thermal equilibrium is a central problem in quantum statistical mechanics. Unlike classical systems, whose thermalization follows from ergodic motion in phase space, the unitary dynamics of closed quantum systems preserves information, challenging the emergence of statistical ensembles. A widely accepted resolution is provided by the eigenstate thermalization hypothesis (ETH), which posits that individual many-body eigenstates already encode thermal behavior for local observables~\cite{PhysRevE.50.888,Srednicki_1999,PhysRevA.43.2046,Deutsch:2018ulr,2008Natur.452..854R,PhysRevLett.54.1879,DAlessio:2015qtq}. Specifically, for a few-body operator $O$ in the energy eigenbasis $\{|E_i\rangle\}$, ETH asserts
\begin{equation}\label{eq:standard_ETH}
O_{ij} = O(E_i)\, \delta_{ij} + e^{-S(E^+)/2} F^{(2)}_{E^+}(\omega_{ij}) R_{ij},
\end{equation}
where $E^+ = (E_i + E_j)/2$, $\omega_{ij} = E_i - E_j$, $O(E)$ gives the thermal expectation value, $S(E)$ is the thermodynamic entropy, $F^{(2)}_{E}(\omega)$ is a smooth function decaying at large $\omega$, and $R_{ij}$ is a Gaussian random variable with zero mean and unit variance. The diagonal term ensures that long-time averages reproduce thermal predictions, while the exponentially small off-diagonal fluctuations drive equilibration under generic dynamics~\cite{Steinigeweg_2013,DAlessio:2015qtq,Beugeling_2015,Mori:2017qhg,Deutsch:2018ulr,Nation_2018,LeBlond_2020,Noh_2021,Alishahiha:2025rdg}. Experimental demonstrations of quantum thermalization have also been performed in optical lattices~\cite{Kaufman:2016mif,Islam:2015mom}. Subsequently, understanding mechanisms that prevent quantum thermalization has become an important topic, highlighting phenomena such as integrability~\cite{Rigol_2007,Iucci_2009,Polkovnikov_2011,Calabrese_2011,Steinigeweg_2013,Caux_2013,Alba_2015,Vidmar_2016,Brenes_2020,LeBlond_2020,rottoli2025eigenstatethermalizationhypothesiseth}, many-body localization~\cite{RevModPhys.91.021001,2018CRPhy..19..498A,2015ARCMP...6..383A,2015ARCMP...6...15N,Schreiber_2015,Chandran_2015,Ponte_2015,Rademaker_2016,Smith_2016,Imbrie_2016,PhysRevLett.111.127201,PhysRevB.90.174202,PhysRevLett.110.067204,Morong_2021}, and nonunitary dynamics generated by repeated measurements~\cite{Patil_2015,PhysRevB.99.224307,PhysRevB.98.205136,PhysRevX.9.031009,Choi_2020,Tang_2020,gherardini2021thermalizationprocessesinducedquantum,Walls_2024,feng2025manybodyantizenothermalizationzeno}.

Nevertheless, the traditional ETH ansatz fails to capture nontrivial higher-order correlation functions, since all cumulants vanish for Gaussian random variables. To overcome this limitation, the full ETH was introduced~\cite{PhysRevE.99.042139}, in which $R_{ij}$ is treated as a more general random variable. Rather than specifying the underlying distribution of $R_{ij}$, the full ETH is formulated directly in terms of averaged products of matrix elements:
\begin{equation}\label{eq:matrixelement1}
\begin{aligned}
&\overline{O_{i_1i_2}O_{i_2i_3}\cdots O_{i_qi_1}}=e^{-(q-1)S(E^+)} F^{(q)}_{E^+}(\bm{\omega}).
\end{aligned}
\end{equation}
Here, we assume $i_1\neq i_2 \cdots \neq i_q$, introduce the relative energies $\bm{\omega}=(\omega_{i_1i_2},\omega_{i_2i_3},\cdots \omega_{i_{q-1}i_q})$ with $\omega_{i_mi_n}=E_{i_m}-E_{i_n}$, and extend the definition of the center-of-mass energy to $E^+=\sum_{m=1}^qE_{i_m}/q$. The statistical average may be understood as an average over individual energy levels within a narrow energy window. In addition, the full ETH assumes
\begin{equation}\label{eq:matrixelement2}
\begin{aligned}
&\overline{O_{i_1i_2}\cdots O_{i_{k-1}i_1}O_{i_1i_{k+1}}\cdots O_{i_{q}i_1}}\\&\ \ \ \ \ \ \ \ \ =\overline{O_{i_1i_2}\cdots O_{i_{k-1}i_1}}~\overline{O_{i_1i_{k+1}}\cdots O_{i_{q}i_1}}.
\end{aligned}
\end{equation} 
Later developments further highlight the connection between full ETH and free probability~\cite{PhysRevLett.129.170603,PhysRevLett.134.140404,Jindal_2024,PhysRevX.15.011031,camargo2025quantumsignatureschaosfree,fritzsch2025freeprobabilityminimalquantum,Fritzsch_2025,vallini2025refinementseigenstatethermalizationhypothesis}, which identifies free cumulants as the fundamental ingredients governing higher-order correlation functions. Analytical and numerical studies of full ETH and its violations have been carried out in Refs.~\cite{PhysRevLett.134.140404,Pathak:2025sys,Alves:2025jzl,Fritzsch:2025ban,vallini2025refinementseigenstatethermalizationhypothesis}. In particular, the factorization correction for non-integrable Floquet systems is discussed in \cite{vallini2025refinementseigenstatethermalizationhypothesis}. However, a careful finite-size scaling analysis of the corrections to these relations in energy conserved systems remains lacking.

In this work, we perform a systematic finite-size scaling analysis of the full ETH in concrete quantum spin chains, considering both single-site and two-site observables. Following Ref.~\cite{PhysRevLett.134.140404}, we focus on finite-size corrections to correlation functions in the canonical ensemble, as elaborated below, rather than working directly with the matrix-element relations in Eqs.~\eqref{eq:matrixelement1} and \eqref{eq:matrixelement2}. We decompose the finite-size corrections of correlation functions into distinct contributions, including those arising from energy fluctuations, which decay polynomially with system size, and those originating from fluctuations within each energy window, which decay exponentially with system size. In particular, the contribution from fluctuations within each energy window can be predicted using thermal density matrices. This decomposition also clearly demonstrates the validity of the full ETH in chaotic spin chains, despite the fact that corrections to certain correlation functions--those involving multiple summations with different scalings--can exhibit anomalous growth with increasing system size at moderate system sizes. Our results provide a refined understanding of the full ETH and a practical methodology for validating it using exact diagonalization.

\section{Full Eigenstate Thermalization for Correlation Functions}
We now describe our setup by revisiting the predictions of the full ETH for correlation functions~\cite{PhysRevE.99.042139,PhysRevLett.129.170603}. We primarily focus on multitime correlation functions of an operator $O$ at infinite temperature, which are defined as
\begin{equation}
C^{(q)}(t_1,t_2,\cdots,t_{q-1})\equiv \langle O(t_1)\cdots O(t_{q-1})O(0)\rangle.
\end{equation}
Here, the expectation value is taken over the infinite-temperature ensemble with density matrix $\rho=\mathbb{1}/D$, where $D$ is the Hilbert space dimension. The operator is evolved under Heisenberg evolution $O(t)\equiv e^{iHt} O e^{-iHt}$ with Hamiltonian $H$. 

\subsection{Two-point function}
To see how the correlation function is related to the full ETH ansatz \eqref{eq:matrixelement1} and \eqref{eq:matrixelement2}, we first consider the case $q=2$, which reads $C^{(2)}(t)=D^{-1}\sum_{ij}e^{i\omega_{ij}t}O_{i j} O_{ji}$~\footnote{In fact, the calculation of the two-point function is already captured by the traditional ETH ansatz \eqref{eq:standard_ETH}.}. To employ the full ETH ansatz \eqref{eq:matrixelement1} and \eqref{eq:matrixelement2}, we separate the summation into contributions with $j=i$ and those with $j\neq i$, which gives
\begin{equation}\label{eq:C2}
C^{(2)}(t)=k_2(t)-\frac{1}{D}\sum_{i}O_{i i}^2.
\end{equation}
Here, we introduce the free cumulants $k_q$, which are fundamental objects in free probability theory \cite{PhysRevLett.129.170603}:
\begin{equation}
k_q(\bm{t})=\frac{1}{D}\sum_{i_1\neq i_2\neq \cdots i_q} e^{i\bm{\omega}\cdot \bm{t}}O_{i_1i_2} O_{i_2i_3}\cdots O_{i_{q}i_1},
\end{equation}
with $\bm{t}=(t_1,t_2,\cdots, t_{q-1})$. In the second term of \eqref{eq:C2}, the summation over individual eigenstates within each narrow energy window effectively plays the role of the averaging in \eqref{eq:matrixelement1} and \eqref{eq:matrixelement2}, which identifies $\frac{1}{D}\sum_{i} O_{ii}^2 = \frac{1}{D}\sum_{i} \overline{O_{ii}^2}$. To proceed, there are two key steps:
\begin{enumerate}
\item We apply the factorization in \eqref{eq:matrixelement2}, which yields the approximation $\overline{O_{ii}^2} \approx \overline{O_{ii}}^2$.

\item We note that the summation over $i$ effectively fixes the energy $E_i$ near that of the infinite-temperature ensemble under the saddle-point approximation. The same energy fixing can be implemented by
\begin{equation}
\frac{1}{D}\sum_{i}\overline{O_{ii}}^2\approx \Big(\frac{1}{D}\sum_{i}\overline{O_{ii}}\Big)^2=k_1^2.
\end{equation} 
\end{enumerate}
Putting all ingredients together, we find the ETH prediction $C^{(2)}(t)\approx k_2(t)-k_1^2\equiv C^{(2)}_{\text{ETH}}(t)$. 

  \begin{figure}[t]
    \centering
    \includegraphics[width=0.8\linewidth]{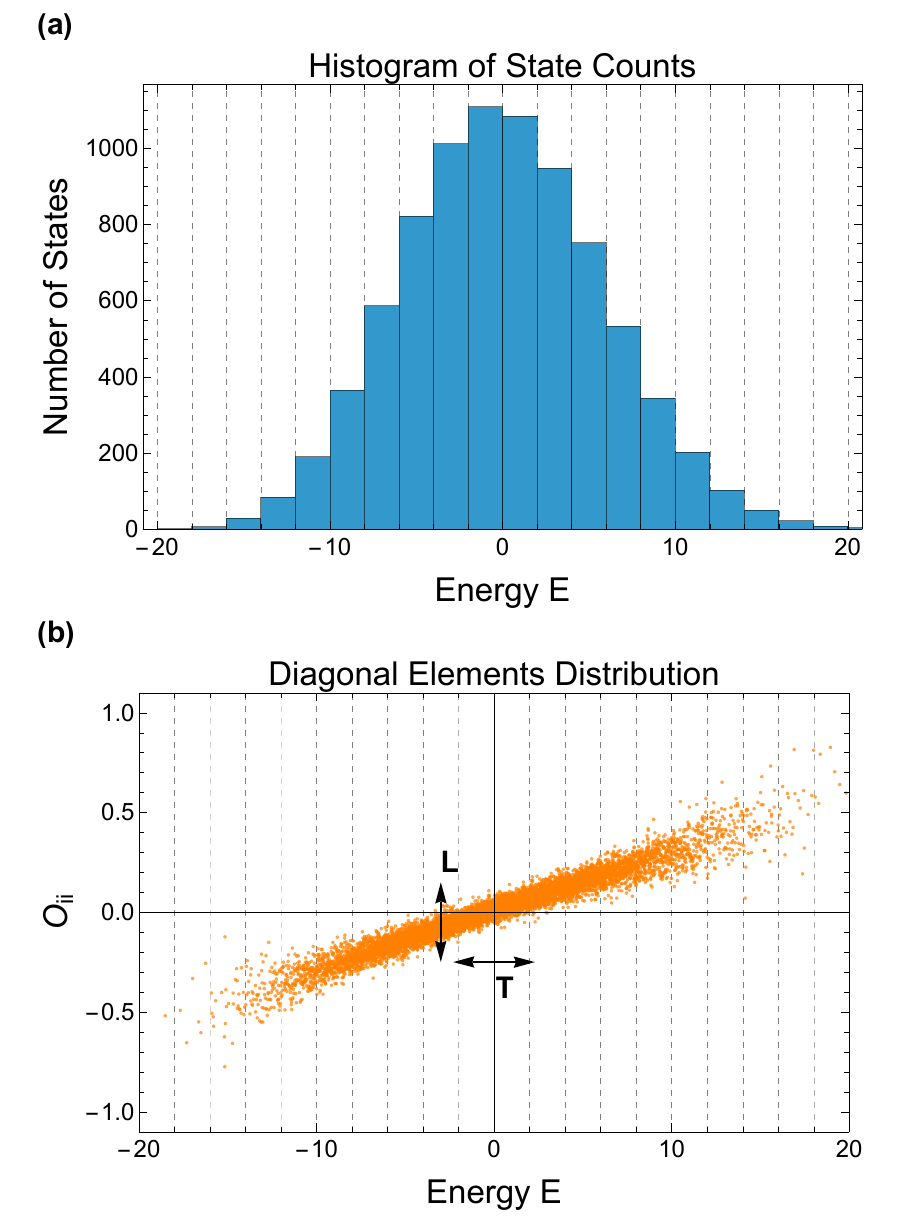}
    \caption{An illustration of the decomposition of finite-size corrections introduced in this work. In panel (a), we introduce a series of narrow energy windows with a fixed width $\Delta=2$ that is independent of the system size. In panel (b), we illustrate the distribution of $O_{ii}$ over all eigenstates, where the total variance consists of the variance within each energy window (longitudinal direction, denoted as L) and the variance between different energy windows (transverse direction, denoted as T).    }
    \label{fig:illustration}
  \end{figure}

The above analysis provides a natural decomposition of the finite-size corrections. We introduce the error of the two-point function as 
\begin{equation}
F_{11}\equiv C^{(2)}_{\text{ETH}}(t)-C^{(2)}(t)=\frac{1}{D}\sum_{i}O_{i i}^2-\Big(\frac{1}{D}\sum_{i}O_{i i}\Big)^2,
\end{equation}
which measures the variance of all diagonal elements $O_{ii}$ across the full spectrum of eigenstates. In step 1, we neglect fluctuations of $O_{ii}$ across individual energy levels within each small energy window, which are expected to be exponentially small in the system size $L$ \cite{PhysRevE.89.042112,DAlessio:2015qtq}. In step 2, we neglect subleading corrections arising from the saddle-point approximation, which originate from energy fluctuations with $\delta E / E \sim O(1/\sqrt{L})$ \cite{landau2013statistical,Touchette_2009}. Therefore, we expect the corresponding contribution to scale polynomially with system size, as explicitly analyzed below. Motivated by these considerations, we introduce a decomposition into longitudinal and transverse variances, $F_{11}=F_{11}^{\text{L}}+F_{11}^{\text{T}}$, where
\begin{equation}\label{eq:decom_c2}
\begin{aligned}
F_{11}^{\text{L}}&\equiv \sum_{E}P_E \bigg[\frac{1}{d_E}\sum_{i\in E}O_{i i}^2-\Big(\frac{1}{d_E}\sum_{i\in E}O_{i i}\Big)^2\bigg],\\
F_{11}^{\text{T}}&\equiv \sum_{E}P_E \bigg[\Big(\frac{1}{d_E}\sum_{i\in E}O_{i i}\Big)^2\bigg]-\bigg[\sum_EP_E\Big(\frac{1}{d_E}\sum_{i\in E}O_{i i}\Big)\bigg]^2.
\end{aligned}
\end{equation}
Here, we introduce a series of narrow energy windows for the eigenenergies of the Hamiltonian $H$ with width $\Delta$, as sketched in FIG.~\ref{fig:illustration}. We denote the number of eigenstates in the window centered around $E$ by $d_E$, and define $P_E = d_E / D$ as a normalized probability distribution. By definition, $F_{11}^{\text{L}}$ measures the variance of $O_{ii}$ within each energy window, weighted by the corresponding probability. In contrast, $F_{11}^{\text{T}}$ measures the variance between different energy windows.

While the analysis of $F_{11}^{\text{L}}$ requires numerical simulations, the scaling of $F_{11}^{\text{T}}$ in the thermodynamic limit can be computed using thermal ensembles. This is because the average within each energy window is equivalent to a microcanonical ensemble average, and the number of states scales as $d_E \sim e^{S(E)}$. Therefore, we have 
\begin{equation}
F_{11}^{\text{T}}=\frac{1}{D}\int dE~e^{S(E)}[O(E)-O(0)]^2.
\end{equation}
Here, $O(E)$ denotes the expectation value of the operator $O$ in the microcanonical ensemble at energy $E$. We assume that the infinite-temperature ensemble corresponds to energy $E=0$, which can always be achieved by shifting the Hamiltonian $H$ by a constant. Since $S(E)$ has a maximum near $E=0$, performing a Taylor expansion yields
\begin{equation}
F_{11}^{\text{T}}\approx\int dE~e^{-\frac{ E^2}{2C_0}}[\partial_E O(0)]^2 E^2\approx {C_0}{[\partial_E O(0)]^2}.
\end{equation}
Here, $C_0 = C_V / \beta^2 \propto L$ is extensive, where $C_V$ is the specific heat in the infinite-temperature limit. We have used the fact that $e^{S(0)} = D$. The derivative $\partial_E O(0)$ scales as $1/L$, since $O(E) \sim O(1)$ for local observables and the energy is extensive. As a consequence, we conclude that $F_{11}^{\text{T}} \propto L^{-1}$ when $\partial_E O(0) \neq 0$. For systems with $\partial_E O(0) = 0$, higher-order expansions must be considered, which lead to a smaller $F_{11}^{\text{T}}$. Nevertheless, we expect that the result still scales polynomially with the system size $L$.

\subsection{Higher-order correlation function}
Next, we generalize the above discussion to higher-order correlation functions. In this work, we focus on the cases $q=3$ and $q=4$. For $q=3$, the full ETH predicts the relation \cite{PhysRevE.99.042139,PhysRevLett.129.170603}
\begin{equation}
\begin{aligned}
C^{(3)}& (t_1,t_2)\approx k_3(t_1,t_2)+[k_2(t_1-t_2)\\
&+k_2(t_1)+k_2(t_1)]k_1+k_1^3\equiv C^{(3)}_{\text{ETH}}(t_1,t_2).
\end{aligned}
\end{equation}
The approximation arises from two contributions: (1) replacing the summation $D^{-1}\sum_i O_{ii}^3$ with $k_1^3=(D^{-1}\sum_i O_{ii})^3$, and (2) replacing the summation $D^{-1}\sum_{i\neq j} e^{i\omega_{ij}t}|O_{ij}|^2 O_{ii}$ with
$$k_2(t)k_1=\frac{1}{D}\sum_{i\neq j} e^{i\omega_{ij}t}|O_{ij}|^2\times \frac{1}{D}\sum_kO_{kk}.$$ Since correlation functions generally decay with time $t$, we introduce error terms evaluated at $t=0$ to benchmark the finite-size corrections as 
\begin{equation}
\begin{aligned}
F_{111}&=\frac{1}{D}\sum_{i}O_{i i}^3-\Big(\frac{1}{D}\sum_{i}O_{i i}\Big)^3,\\
F_{21}&=\frac{1}{D}\sum_{i\neq j}|O_{ij}|^2 O_{ii}-k_2(0)k_1.
\end{aligned}
\end{equation}
For $F_{111}$, it measures how the third-order moment deviates from the expectation value. We can perform a decomposition into longitudinal and transverse components similar to \eqref{eq:decom_c2}, writing $F_{111} = F_{111}^{\text{L}} + F_{111}^{\text{T}}$, where 
\begin{equation}\label{eq:decom_F111}
\begin{aligned}
F_{111}^{\text{L}}&\equiv \sum_{E}P_E \bigg[\frac{1}{d_E}\sum_{i\in E}O_{i i}^3-\Big(\frac{1}{d_E}\sum_{i\in E}O_{i i}\Big)^3\bigg],\\
F_{111}^{\text{T}}&\equiv \sum_{E}P_E \bigg[\Big(\frac{1}{d_E}\sum_{i\in E}O_{i i}\Big)^3\bigg]-\bigg[\sum_EP_E\Big(\frac{1}{d_E}\sum_{i\in E}O_{i i}\Big)\bigg]^3.
\end{aligned}
\end{equation}
Similar to the analysis in previous subsection, $F_{111}^{\text{T}}$ can also be expressed using thermodynamical quantites, and a direct scaling analysis gives $F_{111}^{\text{T}}\lesssim O(L^{-2})$ by expanding $S(E)$ to the order of $E^3$. For $F_{21}$, it is not an independent quantity for operators satisfying $O^2=\mathbb{1}$, which we focus on from now on. Indeed, we have $F_{21}=-m_3+m_1 m_2$, where $m_k=D^{-1}\sum_i O_{ii}^k$. In particular, for Pauli operators with $m_1=0$, this reduces to $F_{21}=-m_3=-F_{111}$. 

Finally, the finite-size corrections for $C^{(4)}(t_1,t_2,t_3)$ can be introduced analogously. Again, we focus only on the error terms at $t_1=t_2=t_3=0$, which include six additional quantities. Firstly, we have an error term
\begin{align}
F_{1111}&=\frac{1}{D}\sum_{i}O_{i i}^4-\Big(\frac{1}{D}\sum_{i}O_{i i}\Big)^4=m_4-m_1^4,
\end{align}
We take $F_{1111}$ as an independent contribution to be analyzed, whose transverse component, defined analoguous to \eqref{eq:decom_c2} and \eqref{eq:decom_F111}, is expected to scale as $L^{-2}$:
\begin{equation}
F_{1111}^{\text{T}}\approx\int dE~e^{-\frac{ E^2}{2C_0}}[\partial_E O(0)]^4 E^4\approx 3{C_0^2}{[\partial_E O(0)]^4}.
\end{equation}
Secondly, we have a new type of error, known as crossing diagram~\cite{PhysRevE.99.042139,PhysRevLett.129.170603,PhysRevLett.134.140404}, defined as $C=D^{-1}\sum_{i\neq j}|O_{ij}|^4$. It is argued that $C$ decays exponentially with system size $L$~\cite{PhysRevE.99.042139,PhysRevLett.129.170603,PhysRevLett.134.140404}. In addition, we have other four error terms:
\begin{align}\label{eq:relation}
F_{22}&=\frac{1}{D}\sum_{i\neq j\neq k}O_{ij}O_{ji}O_{ik}O_{ki}-k_2(0)^2\notag\\&=m_4-m_2^2-C,\\
F_{31}&=\frac{1}{D}\sum_{i\neq j\neq k}O_{ij}O_{jk}O_{ki}O_{ii}-k_3(0,0)k_1\notag\\&=-m_2+2m_4+m_1(m_1+m_2-2m_3)-P,\\
F^{(1)}_{211}&=\frac{1}{D}\sum_{i\neq j}O_{ii}^2O_{ij}O_{ji}-k_2(0)k_1^2\notag\\&=m_2-m_4-m_1^2(1-m_2),\\
F^{(2)}_{211}&=\frac{1}{D}\sum_{i\neq j}O_{ii}O_{jj}O_{ij}O_{ji}-k_2(0)k_1^2\notag\\&=-m_4-m_1^2(1-m_2)+P.
\end{align}
These terms quantify the factorization relation in \eqref{eq:matrixelement2}, which includes loops of indices with $k>1$. For conciseness, we introduce $P=\frac{1}{D}\sum\limits_{ij} O_{ij} O_{ji} O_{ii} O_{jj}$, which is the only new ingredient in addition to $F_{1111}$ and $C$. Theoretically, all error terms should vanish in the thermodynamical limit using the saddle-point analysis~\cite{PhysRevLett.129.170603}. As a result, we expect these terms to decay polynomially with system size. However, the complex structures in these equations make their finite-size scaling extremely difficult to identify without the decomposition into $m_k$, $C$, and $P$, as explicitly demonstrated in the next section. As a result, we will present numerical results for all of these error terms.

\section{Numerical Results}
In this section, we present numerical results for the error terms and their decompositions defined in the previous section. We consider the mixed field Ising model:
\begin{equation}    
\hat{\mathrm{H}}=J\sum\limits_{i=1}^{L}Z_i Z_{i+1}+w\sum\limits_{i=1}^{L}X_i+h\sum\limits_{i=1}^{L}Z_i.
\end{equation}
We choose $J=1$, $w=1.05$, and $h=\frac{\sqrt{5}-1}{2}$ to place the model in the chaotic regime~\cite{Ba_uls_2011,Kim_2013,Atas_2013,atas2015quantumisingmodeltransverse,Kormos_2016,Noh_2021_2,Peng_2022,Chiba_2024,pirmoradian2025investigationquantumchaoslocal}. Focusing on open boundary conditions, the system exhibits reflection symmetry. We therefore restrict our analysis to the $+1$ parity sector and consider only operators that are invariant under the parity transformation. In the Appendix \ref{appendixa}, we present a parallel discussion of the random-field XXZ model, which does not exhibit the reflection symmetry. The results for both models are qualitatively the same.

\begin{figure}[t]
    \centering
    \includegraphics[width=0.99\linewidth]{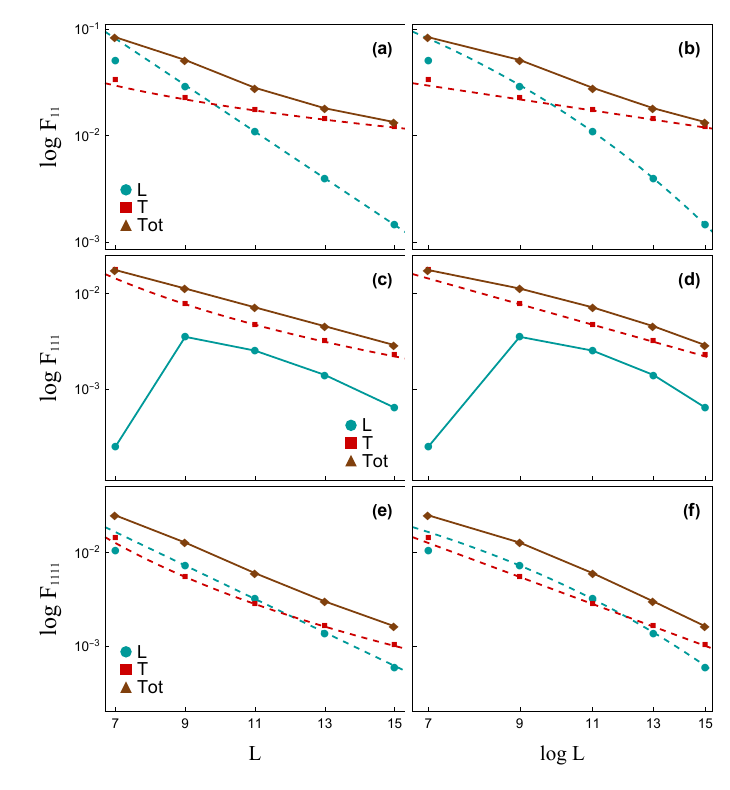}
    \caption{Numerical results for the error terms $F_{11}$, $F_{111}$, and $F_{1111}$ of the mixed-field Ising model as a function of system size $L$ (odd) for the single-site operator $O=Z_{(L+1)/2}$. Here, we fix the width of the energy window to $\Delta=2$. The results are presented in both log plots and log--log plots. The dashed lines correspond to fits to the last four data points, as described in the main text, while the solid lines serve as guides to the eye.}
    \label{fig:Isings1}
\end{figure}

\subsection{Single-site operator}
We now present results for a single-site operator. We focus on systems with odd $L$ and choose the operator to be the Pauli-$Z$ operator at the center, $O=Z_{(L+1)/2}$. The corresponding first moment, $m_1$, remains zero even after fixing the parity sector. Following the previous discussion, we first investigate the error terms $F_{11}=m_2$, $F_{111}=m_3$, and $F_{1111}=m_4$, which correspond to moments of the diagonal matrix elements. The results are presented in FIG.~\ref{fig:Isings1} using both log plots and log--log plots. From the raw data for both $F_{11}$ and $F_{1111}$, we already observe that the longitudinal components decay exponentially with system size, corresponding to straight lines in the log plots. Therefore, we perform fits of the form $a \exp(-b L)$ to these curves using the last four data points, and the results are shown as green dashed lines. For $F_{111}$, the results exhibit a faster-than-exponential decay, as indicated by the increasing slope in the log plot. This behavior originates from cancellations between terms with positive and negative $O_{ii}$. 

\begin{figure}[t]
    \centering
    \includegraphics[width=0.99\linewidth]{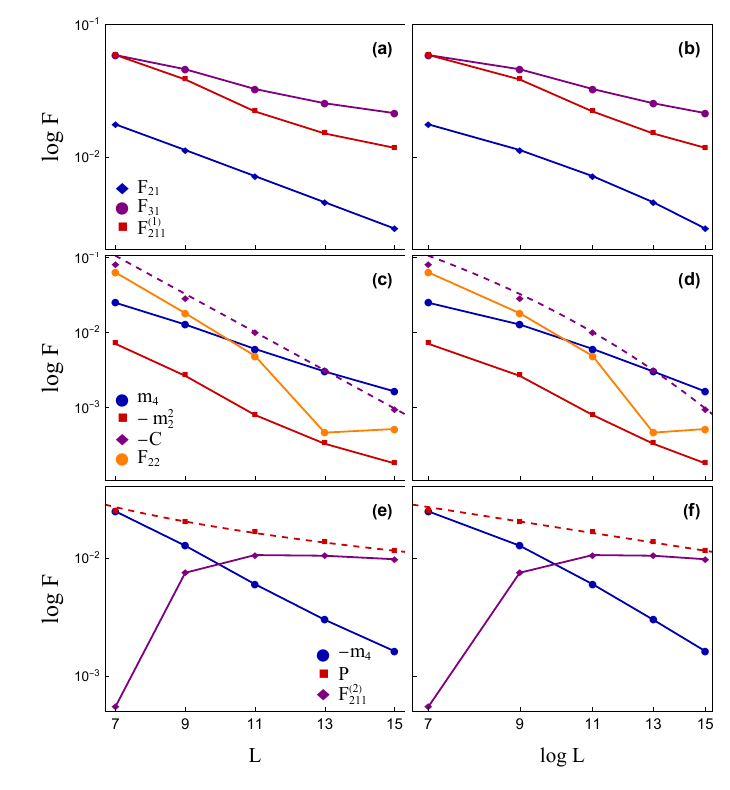}
    \caption{Numerical results for the error terms $F_{21}$, $F_{31}$, $F_{211}^{(1)}$, $F_{22}$, and $F_{211}^{(2)}$ of the mixed-field Ising model as a function of system size $L$ (odd) for the single-site operator $O=Z_{(L+1)/2}$ with $\Delta=2$. The results are presented in both log plots and log--log plots. The dashed lines correspond to fits to the last four data points, as described in the main text, while the solid lines serve as guides to the eye.}
    \label{fig:Isings2}
\end{figure}

On the other hand, the raw data in the transverse direction all exhibit power-law decay, as demonstrated by the log--log plots. We therefore fit the data with $a/L^b$, and the results are shown as red dashed lines. For $F_{11}$, $F_{111}$, and $F_{1111}$, the resulting exponents $b$ are $1.1890$, $2.4640$, and $3.3120$, respectively (see TABLE.~\ref{tab:Isingsscaling}), which include finite-size corrections to the theoretical predictions of $1$, $2$, and $2$ presented in the previous section. Nevertheless, the power-law behavior is already evident despite the moderate system sizes. We further emphasize that the decomposition into longitudinal and transverse components is crucial for clearly validating the finite-size scaling of $F_{11}$, $F_{111}$ and $F_{1111}$. Without this decomposition, the total error term does not exhibit a clear scaling form, since contributions from different components are comparable at the accessible system sizes.

Next, we proceed to other error terms, including $F_{21}$, $F_{31}$, $F_{211}^{(1)}$, $F_{22}$, and $F_{211}^{(2)}$. The results are presented in FIG.~\ref{fig:Isings2}. First, the results for $F_{21}$, $F_{31}$, and $F_{211}^{(1)}$ decay monotonically with system size. In particular, both $F_{21}$ and $F_{211}^{(1)}$ depend only on the moments $m_k$, which have already been analyzed in FIG.~\ref{fig:Isings1}. As a result, we conclude that they all decay polynomially with the system size $L$ in the large-$L$ limit, although this behavior is not evident in finite-size data without the decomposition. The result for $F_{31}$ contains a new ingredient, $P$, which is shown in panels (e-f). We find that $P$ also exhibits power-law decay with system size $L$, and the corresponding fitting results (shown in dashed lines in FIG.~\ref{fig:Isings2} (e-f)) are presented in Table~\ref{tab:Isingsscaling}. This indicates that the dominant contribution to $P$ also originates from energy fluctuations.

\begin{table}[t]
\centering
\caption{Fitted parameters for the single-site operator $O=Z_{(L+1)/2}$ of the mixed-field Ising model.}
\label{tab:Isingsscaling}
\begin{tabular}{l l S[table-format=1.4] S[table-format=1.4]}
\toprule
\textbf{Component} & \textbf{Model Expression} & \textbf{a} & \textbf{b} \\
\midrule
$~~~~~~F_{11}^{\text{L}}$    & $~~~~~a \exp(-b L)$ & 2.7530 & 0.5032 \\
$~~~~~~F_{11}^{\text{T}}$    & $~~~~~a / L^b$      & 0.2975 & 1.1890 \\
\midrule
$~~~~~~F_{111}^{\text{L}}$   & $~~~~~a \exp(-b L)$ & {---}  & {---}  \\
$~~~~~~F_{111}^{\text{T}}$   & $~~~~~a / L^b$      & 1.7390 & 2.4640 \\
\midrule
$~~~~~~F_{1111}^{\text{L}}$  & $~~~~~a \exp(-b L)$ & 0.2895 & 0.4104 \\
$~~~~~~F_{1111}^{\text{T}}$  & $~~~~~a / L^b$      & 7.8840 & 3.3120 \\
\midrule
$~~~~~~C$          & $~~~~~a \exp(-b L)$ & 6.3150 & 0.5861 \\
$~~~~~~P$           & $~~~~~a / L^b$      & 0.2346 & 1.1150 \\
\bottomrule
\end{tabular}
\end{table}

Finally, numerical results for $F_{22}$ and $F_{211}^{(2)}$ are presented in FIG.~\ref{fig:Isings2}(c–f). Surprisingly, these results exhibit anomalous growth as the system size increases to $L=15$, which naively indicates a violation of the full ETH, despite the fact that the quantum spin chain is chaotic. As we mentioned in the previous discussions, the important observation to resolve this puzzle is the existence of multiple terms. For $m_1=0$, the relations in \eqref{eq:relation} becomes
\begin{equation}
F_{22}=m_4-m_2^2-C,\ \ \ \ \ \ F_{211}^{(2)}=-m_4+P.
\end{equation}
If we plot the results for each component individually, they show clear decay as the system size increases. As elaborated in FIG.~\ref{fig:Isings1}, both $m_4$ and $m_2$ exhibit combinations of exponential and power-law decay with system size $L$. The results for $C$ also clearly show exponential decay with $L$, as indicated by the fits shown using the purple dashed lines. However, when these components are combined to obtain $F_{22}$, the result exhibits a kink due to an interchange of the dominant contributions at accessible system sizes. A similar phenomenon occurs for $F_{211}^{(2)}$. Our results highlight the importance to analyze individual contributions for validating the full ETH in quantum lattice systems.

\begin{figure}[t]
    \centering
    \includegraphics[width=0.99\linewidth]{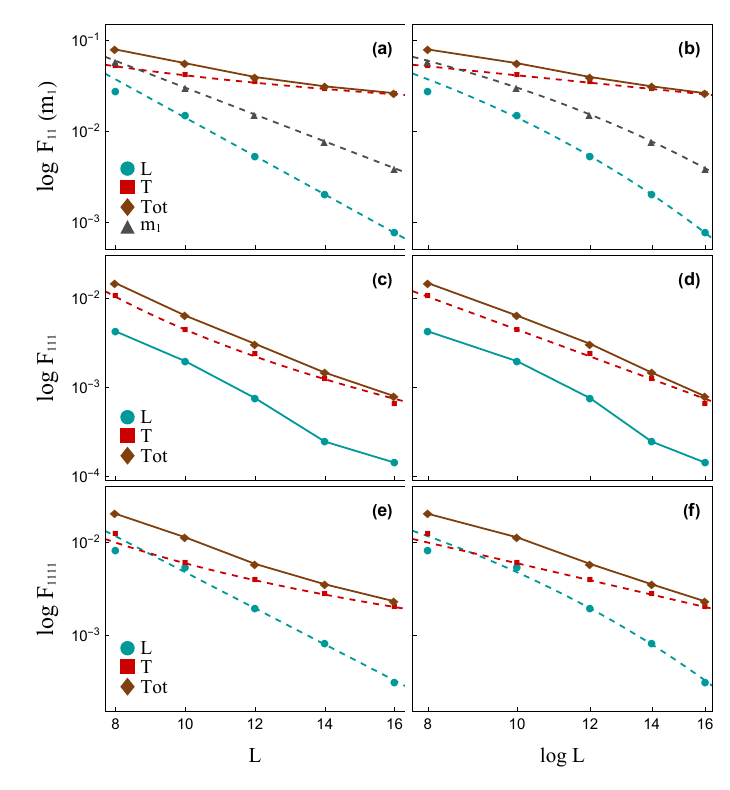}
    \caption{Numerical results for the error terms $F_{11}$, $F_{111}$, and $F_{1111}$ of the mixed-field Ising model as a function of the system size $L$ (even) for the two-site operator $O = Z_{L/2} Z_{L/2+1}$. We fix the width of the energy window to $\Delta=2$. The results are presented in both log plots and log--log plots. The dashed lines correspond to fits to the last four data points, while the solid lines serve as guides to the eye.}
    \label{fig:Isingt1}
\end{figure}

\subsection{Two-site operator}
We next consider a two-site operator for systems with even $L$. We choose the operator to be $O = Z_{L/2} Z_{L/2+1}$. While the first moment $m_1$ vanishes in the full Hilbert space, it becomes nonzero when restricted to states with even parity. To obtain $m_1(L)$, we first focus on the pair of qubits at sites $L/2$ and $L/2+1$. There are two states, $\ket{\uparrow\uparrow}$ and $\ket{\downarrow\downarrow}$, with eigenvalue $Z_{L/2} Z_{L/2+1} = 1$; both states have even parity. There are also two states, $\frac{1}{\sqrt{2}}(\ket{\uparrow\downarrow}+\ket{\downarrow\uparrow})$ and $\frac{1}{\sqrt{2}}(\ket{\uparrow\downarrow}-\ket{\downarrow\uparrow})$, with eigenvalue $-1$, and only one of them has even parity. Therefore, we expect
\begin{equation}
m_1(L)= \frac{D_{\text{even}}(L-2)-D_{\text{odd}}(L-2)}{3D_{\text{even}}(L-2)+D_{\text{odd}}(L-2)}.
\end{equation}
Here, $D_{\text{even}}(L-2)$ and $D_{\text{odd}}(L-2)$ denote the dimensions of the Hilbert spaces with even and odd parity, respectively, for the remaining $L-2$ qubits. To proceed, we perform a similar analysis by adding a pair of qubits to derive the recursion relations for $D_{\text{even}}(L-2)$ and $D_{\text{odd}}(L-2)$, which read
\begin{equation}
\begin{aligned}
D_{\text{even}}(L)&=3D_{\text{even}}(L-2)+D_{\text{odd}}(L-2),\\
D_{\text{odd}}(L)&=3D_{\text{odd}}(L-2)+D_{\text{even}}(L-2).
\end{aligned}
\end{equation}
The solution gives $D_{\text{even/odd}}(L)=2^{L-1}\pm2^{\frac{L}{2}-1}$. As a result, we find $m_1(L)=\frac{1}{2^{L/2}+1}$, which decays exponentially with system size $L$.

\begin{figure}[t]
    \centering
    \includegraphics[width=0.99\linewidth]{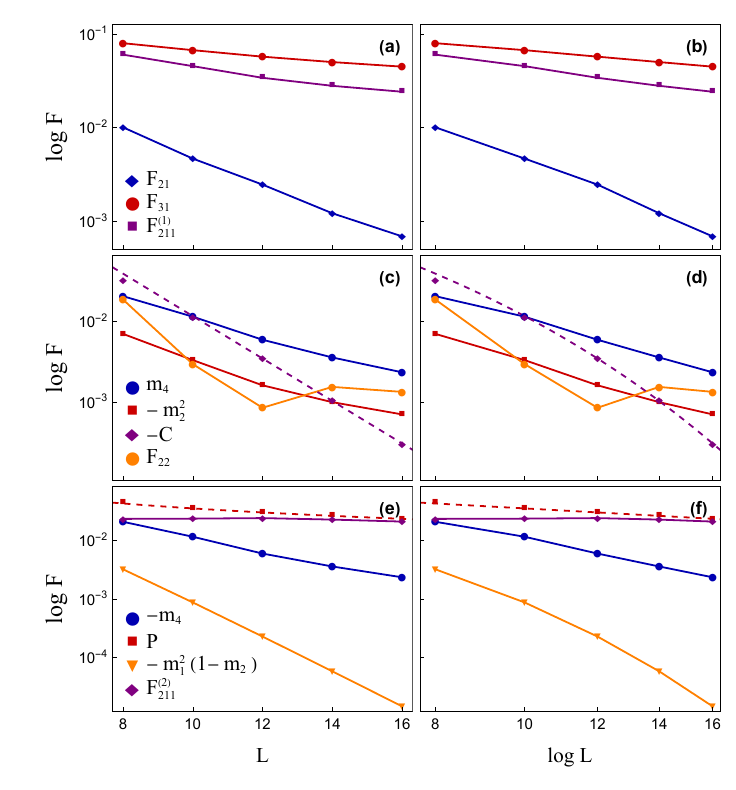}
    \caption{Numerical results for the error terms $F_{21}$, $F_{31}$, $F_{211}^{(1)}$, $F_{22}$, and $F_{211}^{(2)}$ of the mixed-field Ising model as a function of the system size $L$ (even) for the two-site operator $O = Z_{L/2} Z_{L/2+1}$, with $\Delta=2$. The results are presented in both log plots and log--log plots. The dashed lines correspond to fits to the last four data points, while the solid lines serve as guides to the eye.}
    \label{fig:Isingt2}
\end{figure}

\begin{table}[b]
\centering
\caption{Fitted parameters for the two-site operator $O=Z_{L/2} Z_{L/2+1}$ of the mixed-field Ising model.}
\label{tab:Isingtscaling}
\begin{tabular}{l l S[table-format=2.4] S[table-format=1.4]}
\toprule
\textbf{Component} & \textbf{Model Expression} & \textbf{a} & \textbf{b} \\
\midrule
$~~~~~~m_1$          & $~~~~~a \exp(-bL)$      & 0.9132 & 0.3405 \\
\midrule
$~~~~~~F_{11}^{\text{L}}$    & $~~~~~a \exp(-b L)$ & 1.8370 & 0.4865 \\
$~~~~~~F_{11}^{\text{T}}$    & $~~~~~a / L^b$      & 0.4393 & 1.0250 \\
\midrule
$~~~~~~F_{111}^{\text{L}}$   & $~~~~~a \exp(-b L)$ & {---}  & {---}  \\
$~~~~~~F_{111}^{\text{T}}$   & $~~~~~a / L^b$      & 28.1200 & 3.8020 \\
\midrule
$~~~~~~F_{1111}^{\text{L}}$  & $~~~~~a \exp(-b L)$ & 0.4197 & 0.4477 \\
$~~~~~~F_{1111}^{\text{T}}$  & $~~~~~a / L^b$      & 1.2180 & 2.3100 \\
\midrule
$~~~~~~C$          & $~~~~~a \exp(-b L)$ & 4.9650 & 0.6063 \\
$~~~~~~P$           & $~~~~~a / L^b$      & 0.2535 & 0.8670 \\
\bottomrule
\end{tabular}
\end{table}

With this knowledge, we proceed to the numerical results for $F_{11}=m_2-m_1^2$, $F_{111}=m_3-m_1^3$, and $F_{1111}=m_4-m_1^4$. The results, shown in Fig.~\ref{fig:Isingt1}, are analogous to those obtained for the single-site operator in the previous section: the longitudinal component decays exponentially with $L$, while the transverse component decays polynomially. The only exception is $F_{111}^{\text{L}}$. Unlike the single-site case, the slope in the log plot exhibits non-monotonic behavior. Nevertheless, since $|m_3|<m_2$ due to $|O_{ii}|<1$, the decay of $F_{111}^{\text{L}}$ is controlled by that of $F_{11}^{\text{L}}$ and at most polynomial in the system size $L$. The results for the error terms $F_{21}$, $F_{31}$, $F_{211}^{(1)}$, $F_{22}$, and $F_{211}^{(2)}$ are presented in Fig.~\ref{fig:Isingt2}. The error terms $F_{21}$, $F_{31}$, and $F_{211}^{(1)}$ decay monotonically as functions of the system size $L$. In contrast, $F_{22}$ and $F_{211}^{(2)}$ again exhibit anomalous growth in the accessible system sizes. Nevertheless, each individual component shows a clear monotonic decrease as $L$ increases. In particular, the fitted parameters for each components are presented in TABLE.~\ref{tab:Isingtscaling}. This behavior clearly demonstrates the validity of the full ETH for the two-site operator.

  \begin{figure}[t]
    \centering
    \includegraphics[width=0.99\linewidth]{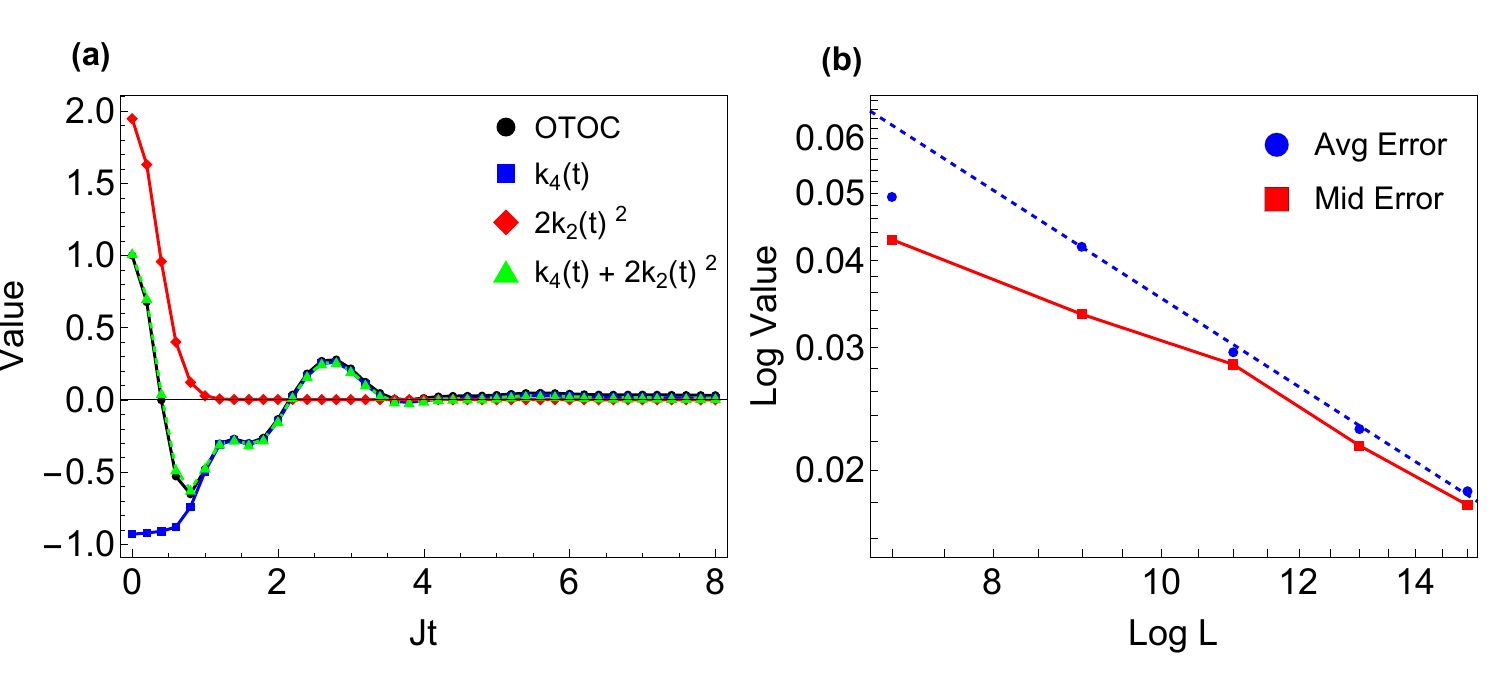}
    \caption{Numerical results for the finite-size corrections of the OTOC for the single-site operator $O = Z_{(L+1)/2}$. In panel (a), we present the numerical results at $L=15$ for the OTOC, the free cumulants, and the prediction from the full ETH. In panel (b), we show the finite-size corrections for both the averaged and mid-time errors.  }
    \label{fig:dynamics}
  \end{figure}

\subsection{Finite-size corrections at finite times }
In the previous discussion, we only considered finite-size corrections at $t=0$. In this subsection, we examine the finite-size scaling of the error at nonzero times. We focus on the single-site operator $O = Z_{(L+1)/2}$ for $q=4$ with $k_1 = m_1 = 0$. We consider the out-of-time-order correlator (OTOC), defined as~\cite{1969JETP...28.1200L,Maldacena:2015waa,Shenker:2014cwa,Roberts:2014isa,Shenker_2014,kitaev2015simple,Hashimoto_2017}
\begin{equation}
\text{OTOC}(t)\equiv C^{(4)}(t,0,t)=\langle O(t)O(0)O(t)O(0)\rangle.
\end{equation}
Using the full ETH, the OTOC can be related to free cumulants as (for $k_1=0$):
\begin{equation}
\text{OTOC}(t)\approx k_4(t,0,t)+2k_2(t)^2\equiv \text{OTOC}_{\text{ETH}}(t).
\end{equation} 
We numerically benchmark the deviation between $\text{OTOC}_{\text{ETH}}(t)$ and $\text{OTOC}(t)$ using two criteria, which capture either the averaged error or the error at a single time. The averaged error is defined as 
\begin{equation}
\mathrm{E}_{\text{avg}}=\left|\frac{1}{T}\int_0^T dt\left[\text{OTOC}_{\text{ETH}}(t)-\mathrm{OTOC}\left(T/2\right)\right]\right|.
\end{equation}
Here, we choose a cutoff time $T$. For the single-time error, we focus on the time $T/2$, which we refer to as the mid-time error, defined as $\mathrm{E}_{\text{Mid}} = |\text{OTOC}_{\text{ETH}}(T/2) - \text{OTOC}(T/2)|$. The numerical results are presented in FIG.~\ref{fig:dynamics} for $T=8$, which suggests that both errors decay polynomially in system size. Using the last four data points, we fit the averaged error with $a/L^b$, obtaining $a \approx 1.4$ and $b \approx 1.6$, as shown by the blue dashed lines. These results further support the validity of the full ETH at finite evolution times $t$.

\section{Discussions}

In summary, we have performed a comprehensive numerical study of finite-size corrections to the full eigenstate thermalization hypothesis in chaotic quantum spin chains. By decomposing error terms into longitudinal components, which capture fluctuations within narrow energy windows, and transverse components, which capture fluctuations across different energy windows, we identified their distinct scaling behaviors. Our results show that longitudinal contributions decay exponentially with system size, while transverse contributions follow power-law scaling consistent with thermodynamic predictions. Importantly, we resolved apparent anomalies in certain higher-order correlation functions, demonstrating that these are finite-size effects arising from the interplay of multiple contributions rather than genuine violations of the full ETH. The decomposition framework and systematic finite-size analysis presented here provide a practical and reliable methodology for validating the full ETH in both single-site and multi-site observables, offering insights into the emergence of thermal behavior in finite quantum systems.

We conclude our work with several remarks. First, a detailed analysis at finite temperatures may reveal additional structures in the finite-size scaling behavior, potentially offering new insights into the thermalization process in quantum many-body systems. Second, it is natural to ask whether energy conservation significantly affects the finite-size scaling; this question could be addressed by studying eigenstates of chaotic quantum circuit models with and without conservation laws. Third, exploring the finite-size scaling of the full ETH in integrable~\cite{Pathak:2025sys} or localized systems may provide a clearer distinction between these systems and chaotic ones, and help identify universal versus nonuniversal features of quantum thermalization. Finally, understanding how the full ETH manifests in conformal field theories with large central charge \cite{Lashkari:2016vgj} could reveal a rich interplay among quantum chaos, holographic duality, and the structure of operator correlations. We leave all these studies to future work.

\vspace{5pt}
\textit{Acknowledgement.}
We thank Yanting Cheng and Ning Sun for helpful discussions. This project is supported by the NSFC under grant 12374477, the Shanghai Rising-Star Program under grant number 24QA2700300, and the Quantum Science and Technology-National Science and Technology Major Project 2024ZD0300101.

\appendix
\section{The random-field XXZ model} \label{appendixa}
In this appendix, we present numerical results for the random-field XXZ model. The Hamiltonian reads 
\begin{equation}    
\hat{\mathrm{H}}=J\sum\limits_{i=1}^{L}\left(X_i X_{i+1}+Y_iY_{i+1}\right)+J_z\sum\limits_{i=1}^{L}Z_iZ_{i+1}+\sum\limits_{i=1}^{L}h_iZ_i.
\end{equation}
We focus on open boundary conditions and choose $J=1$ and $J_z=1.05$. The random fields $h_i$ are independently sampled from the uniform distribution $[-0.75, 0.75]$, a regime in which the system resides in the thermalized phase~\cite{Znidaric2008,pal2010manybodylocalizationtransition,Luitz_2015,Serbyn_2015,RevModPhys.91.021001,_untajs_2020}. Due to the spin rotation symmetry (along the $z$ direction) of the Hamiltonian. we restrict our discussion to the $S_z=0$ sector. We focus on the same single-site and two-site operators as those employed in the mixed-field Ising model, which are invariant under the spin rotation.

\begin{figure}[t]
    \centering
    \includegraphics[width=0.99\linewidth]{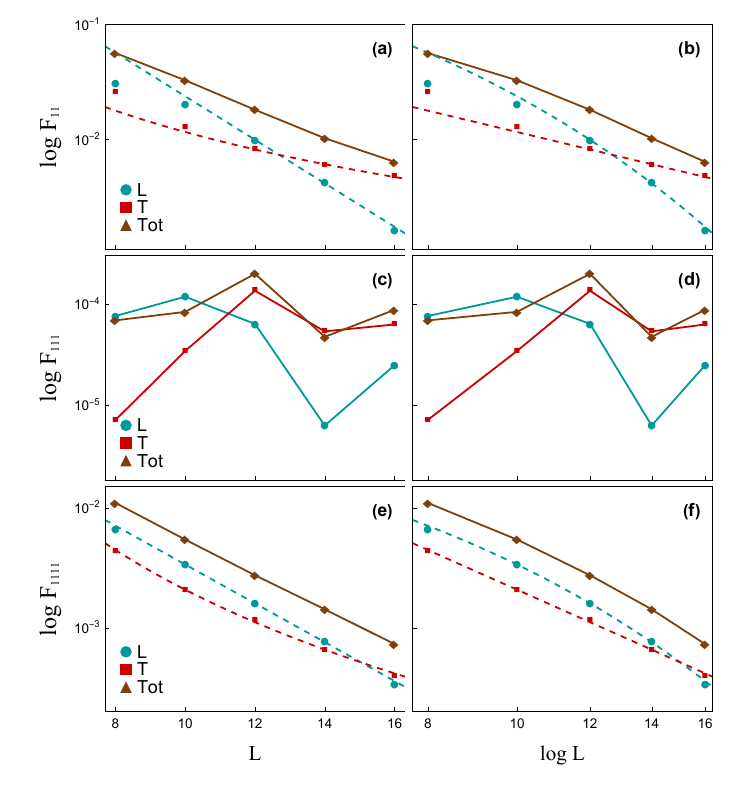}
    \caption{Numerical results for the error terms $F_{11}$, $F_{111}$, and $F_{1111}$ of the random-field XXZ model as a function of system size $L$ for the single-site operator $O=Z_{L/2}$. Here, we fix the width of the energy window to $\Delta=1$. The results are presented in both log plots and log--log plots. The dashed lines correspond to fits to the last four data points, while the solid lines serve as guides to the eye.}
    \label{fig:xxzs1}
\end{figure}

\subsection{Single-site operator} \label{single-site-XXZ}

\begin{figure}[t]
    \centering
    \includegraphics[width=0.99\linewidth]{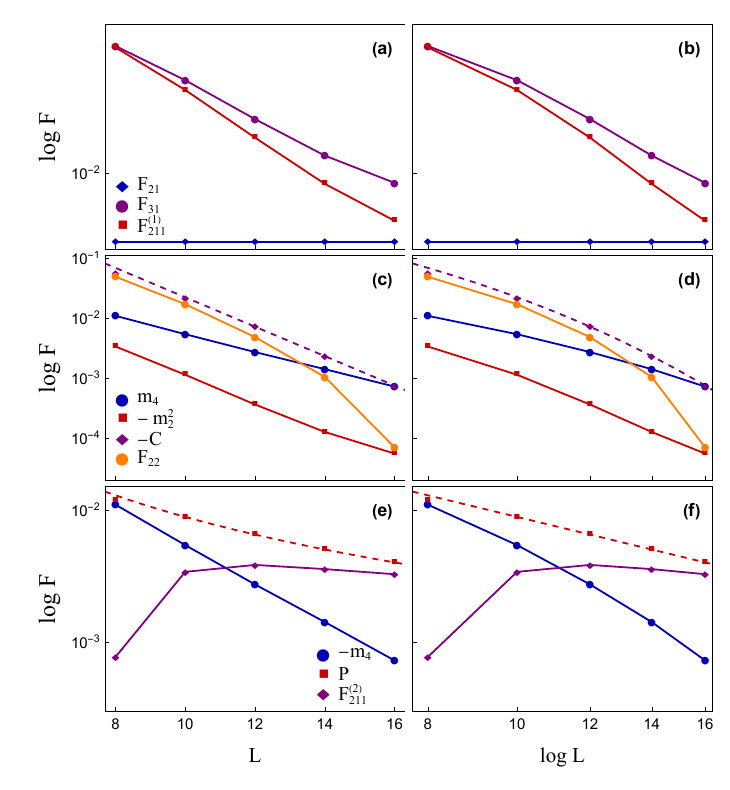}
    \caption{Numerical results for the error terms $F_{21}$, $F_{31}$, $F_{211}^{(1)}$, $F_{22}$, and $F_{211}^{(2)}$ of the random-field XXZ model as a function of system size $L$ for the single-site operator $O=Z_{L/2}$ with $\Delta=1$. The results are presented in both log plots and log--log plots. The dashed lines correspond to fits to the last four data points, while the solid lines serve as guides to the eye.}
    \label{fig:XXZs2}
\end{figure}

Here, we focus on systems with even $L$ and choose the operator $O=Z_{L/2}$. Similar to the mixed-field Ising model, $m_1$ remains zero after fixing the spin sector. The numerical results for the error terms $F_{11}=m_2$, $F_{111}=m_3$, and $F_{1111}=m_4$ are presented in
 FIG.~\ref{fig:xxzs1}. Following the same notation as that for the mixed-field Ising model. The longitudinal and transverse components of $F_{11}$ and $F_{1111}$ decay exponentially and polynomially with $L$, respectively, which is consistent with the previous conclusion. Accordingly, we perform fits of the form $a\exp(-bL)$ for $F_{11}^{\text{L}}$, $F_{1111}^{\text{L}}$, and fits of the form $a/L^b$ for $F_{11}^{\text{T}}$, $F_{1111}^{\text{T}}$, in both cases using the last four data points. We find that the behaviors of both longitudinal and transverse components $F_{111}^{\text{L}}$, $F_{111}^{\text{T}}$ are non-monotonic. Nevertheless, as pointed out in the main text, their behaviors should be controlled by that of $F_{11}^{\text{L}}$ and $F_{11}^{\text{T}}$, remaining at most polynomial in the system size $L$.
The results for the error terms $F_{21}$, $F_{31}$, $F_{211}^{(1)}$, $F_{22}$, and $F_{211}^{(2)}$ are presented in Fig.~\ref{fig:XXZs2}.
Since these terms exhibit behaviors similar to those observed in the mixed-field Ising model, we employ the same decomposition method to analyze $F_{22}$ and $F_{211}^{(2)}$. The fitted parameters for each component are presented in TABLE.~\ref{tab:XXZsscaling}. These results validate the full ETH for the single-site operator.

\begin{table}[b]
\centering
\caption{Fitted parameters for the single-site operator $O=Z_{L/2}$ of the random-field XXZ model.}
\label{tab:XXZsscaling}
\begin{tabular}{l l S[table-format=1.4] S[table-format=1.4]}
\toprule
\textbf{Component} & \textbf{Model Expression} & \textbf{a} & \textbf{b} \\
\midrule
$~~~~~~F_{11}^{\text{L}}$    & $~~~~~a\exp(-b L)$ & 1.8600 & 0.4360 \\
$~~~~~~F_{11}^{\text{T}}$    & $~~~~~a / L^b$      & 0.9643 & 1.9190 \\
\midrule
$~~~~~~F_{111}^{\text{L}}$   & $~~~~~a \exp(-b L)$ & {---}  & {---}  \\
$~~~~~~F_{111}^{\text{T}}$   & $~~~~~a / L^b$      & {---}  & {---} \\
\midrule
$~~~~~~F_{1111}^{\text{L}}$  & $~~~~~a \exp(-b L)$ & 0.1394 & 0.3725 \\
$~~~~~~F_{1111}^{\text{T}}$  & $~~~~~a / L^b$      & 5.4190  & 3.4170 \\
\midrule
$~~~~~~C$          & $~~~~~a \exp(-b L)$ & 6.2120 & 0.5630 \\
$~~~~~~P$           & $~~~~~a / L^b$      & 0.4270 & 1.6830 \\
\bottomrule
\end{tabular}
\end{table}

\begin{figure}[t]
    \centering
    \includegraphics[width=0.99\linewidth]{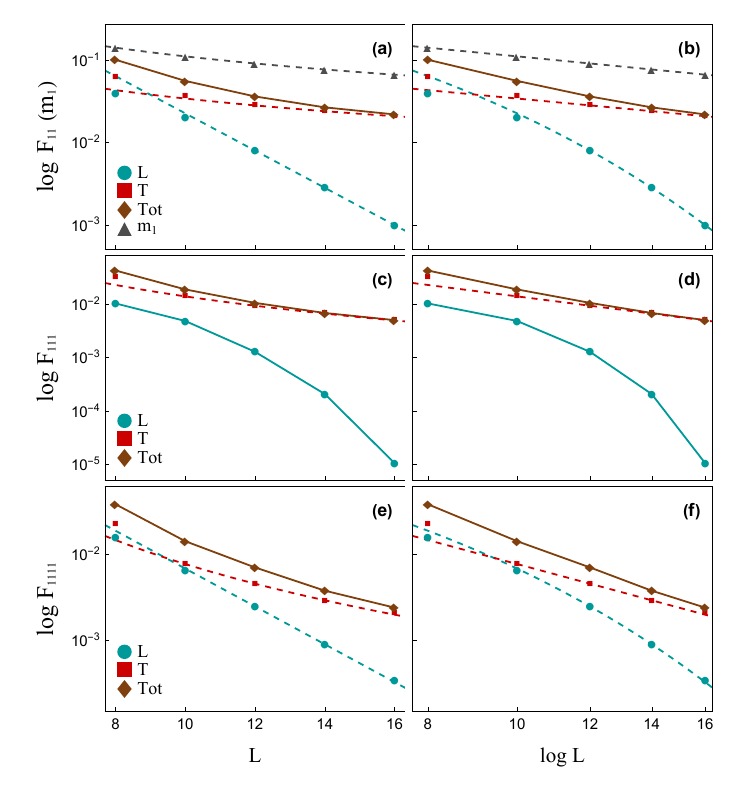}
    \caption{Numerical results for the error terms $F_{11}$, $F_{111}$, and $F_{1111}$ of the random-field XXZ model as a function of system size $L$ for the two-site operator $O=Z_{L/2}Z_{L/2+1}$. Here, we fix the width of the energy window to $\Delta=1$. The results are presented in both log plots and log--log plots. The dashed lines correspond to fits to the last four data points, while the solid lines serve as guides to the eye.}
    \label{fig:XXZt1}
\end{figure}

\subsection{Two-site operator} \label{two-site-XXZ}

Next, we focus on systems with even $L$ and choose the operator $O=Z_{L/2}Z_{L/2+1}$. Unlike the previous case, $m_1$ becomes nonzero in the $S_z=0$ sector for this operator, similar to the observations in the mixed-field Ising model. However, the distinct symmetry sectors imply different finite-size scaling behaviors for $m_1(L)$. To determine $m_1(L)$, we first consider the pair of qubits at sites $L/2$ and $L/2+1$. Within this subspace, there are two states, $\ket{\uparrow\uparrow}$ and $\ket{\downarrow\downarrow}$, with the eigenvalue $Z_{L/2} Z_{L/2+1} = 1$, and two states, $\ket{\uparrow\downarrow}$ and $\ket{\downarrow\uparrow}$, with the eigenvalue $-1$. Therefore, we expect
\begin{equation}
m_1(L)= 2\frac{\binom{L-2}{\frac{L}{2}-2}-\binom{L-2}{\frac{L}{2}-1}}{\binom{L}{L/2}}=-\frac{2}{L-1}\binom{n}{k}\frac{\binom{L-1}{\frac{L}{2}-1}}{\binom{L}{\frac{L}{2}}}\sim-\frac{1}{L}.
\end{equation}
In the second step, we invoke the identity $\binom{n}{k}-\binom{n}{k-1}=\frac{n-2k+1}{n+1}\binom{n+1}{k}$. For the final step, we employ Stirling's approximation, $n!\sim\sqrt{2\pi n}\left(\frac{n}{e}\right)^n$, to obtain the asymptotic behavior.

\begin{figure}[t]
    \centering
    \includegraphics[width=0.99\linewidth]{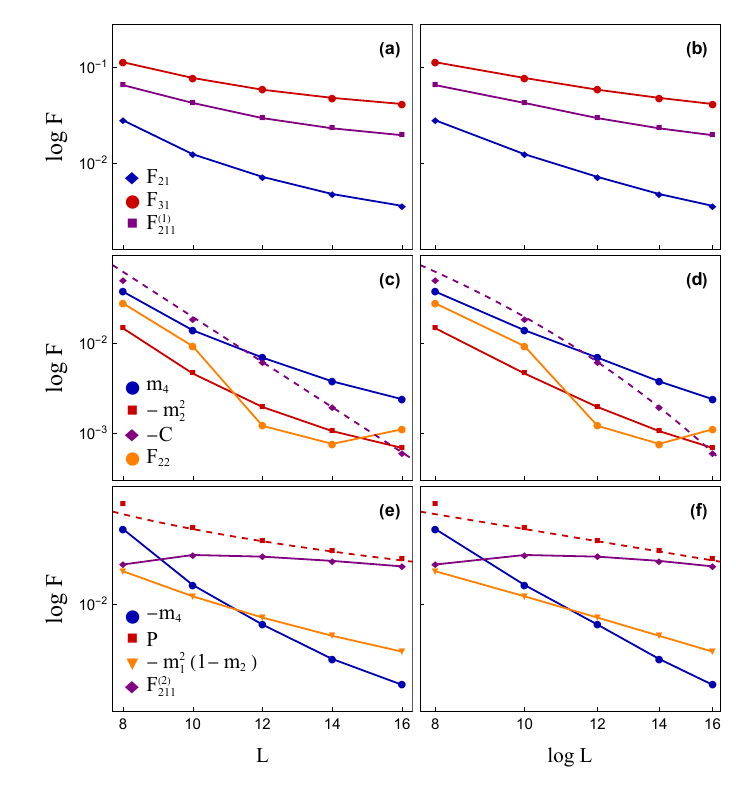}
    \caption{Numerical results for the error terms $F_{21}$, $F_{31}$, $F_{211}^{(1)}$, $F_{22}$, and $F_{211}^{(2)}$ of the random-field XXZ model as a function of system size $L$ for the two-site operator $O=Z_{L/2}Z_{L/2+1}$ with $\Delta=1$. The results are presented in both log plots and log--log plots. The dashed lines correspond to fits to the last four data points, while the solid lines serve as guides to the eye.}
    \label{fig:XXZt2}
\end{figure}

We then proceed to the numerical results for $F_{11}=m_2-m_1^2$, $F_{111}=m_3-m_1^3$, $F_{1111}=m_4-m_1^4$, as displayed in Fig.~\ref{fig:XXZt1}. 
The results for additional error terms, including $F_{21}$, $F_{31}$, $F_{211}^{(1)}$, $F_{22}$, and $F_{211}^{(2)}$ are presented in Fig.~\ref{fig:XXZt2}.
Their scaling behaviors are analogous to those observed for the single-site operators in the random-field Ising model; in particular, $F_{111}$ exhibit a faster-than-exponential decay, similar to that for the single-site case in the mixed-field Ising model. The fitted parameters for each component are presented in TABLE.~\ref{tab:XXZtscaling}. In summary, we have presented numerical results for both single-site and two-site operators in the thermalized phase of the random-field XXZ model. Through the decomposition framework discussed in the main text, we conclude that both cases validate the full ETH.

\begin{table}[t]
\centering
\caption{Fitted parameters for the two-site operator $O=Z_{L/2} Z_{L/2+1}$ of the random-field XXZ model.}
\label{tab:XXZtscaling}
\begin{tabular}{l l S[table-format=1.4] S[table-format=1.4]}
\toprule
\textbf{Component} & \textbf{Model Expression} & \textbf{a} & \textbf{b} \\
\midrule
$~~~~~~m_1$          & $~~~~~a / L^b$      & 1.3630 & 1.0890 \\
\midrule
$~~~~~~F_{11}^{\text{L}}$    & $~~~~~a \exp(-b L)$ & 4.1150 & 0.5206 \\
$~~~~~~F_{11}^{\text{T}}$    & $~~~~~a / L^b$      & 0.3830 & 1.0500 \\
\midrule
$~~~~~~F_{111}^{\text{L}}$   & $~~~~~a \exp(-b L)$ & {---}  & {---}  \\
$~~~~~~F_{111}^{\text{T}}$   & $~~~~~a / L^b$      & 2.2090 & 2.2030 \\
\midrule
$~~~~~~F_{1111}^{\text{L}}$  & $~~~~~a \exp(-b L)$ & 1.0440 & 0.5039 \\
$~~~~~~F_{1111}^{\text{T}}$  & $~~~~~a / L^b$      & 5.6220 & 2.8680 \\
\midrule
$~~~~~~C$          & $~~~~~a \exp(-b L)$ & 6.6600 & 0.5809 \\
$~~~~~~P$           & $~~~~~a / L^b$      & 0.5738 & 1.1810 \\
\bottomrule
\end{tabular}
\end{table}

\section{The level spacing ratio} \label{appendixb}
In this section, we provide the level spacing ratio $\scalebox{1.2}{\text{r}}$ for the models discussed in the main text and the preceding section. The ratio $\scalebox{1.2}{\text{r}}$ serves as a standard diagnostic tool for identifying quantum chaos and distinguishing it from integrability.
\begin{equation}
    \scalebox{1.2}{\text{r}}_n=\frac{\text{min}\{s_n,s_{n+1}\}}{\text{max}\{s_n,s_{n+1}\}}.
\end{equation}

\begin{figure}[h]
    \centering
    \includegraphics[width=0.85\linewidth]{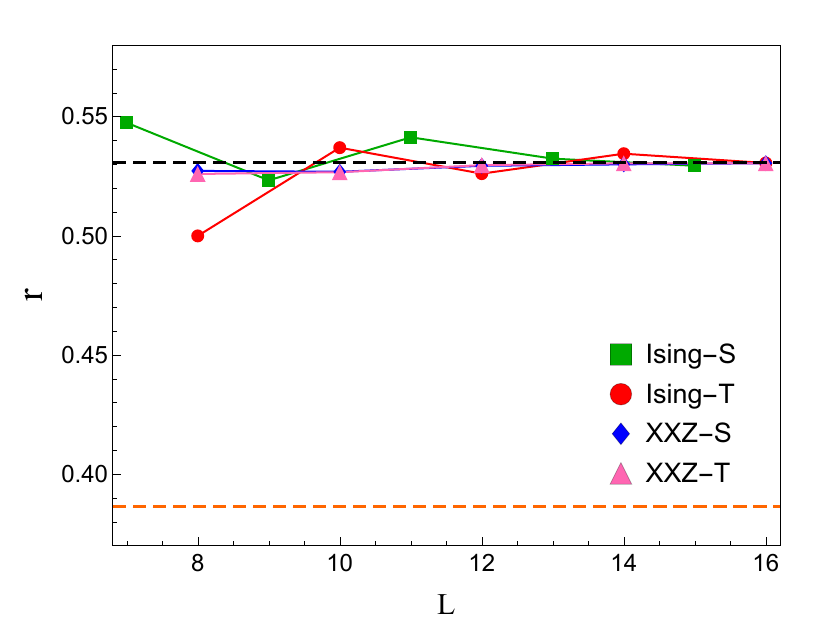}
    \caption{Numerical results for level spacing ratios \scalebox{1.2}{\text{r}} in the mixed-field Ising and random-field XXZ models, considering both single-site and two-site operators. The black and orange dashed lines indicate the theoretical values for Wigner-Dyson and Poisson statistics, respectively. The solid lines serve as guides to the eye.}
    \label{fig:levelspacing}
\end{figure}

Here, $s_n=E_{n+1}-E_n$ denotes the spacing between adjacent energy levels, and $\scalebox{1.2}{\text{r}}$=$\frac{1}{D}\sum_{n}\scalebox{1.2}{\text{r}}_n$ is the average level spacing ratio. For chaotic systems obeying Wigner-Dyson statistics, the expected value is $\scalebox{1.2}{\text{r}}_{\scriptscriptstyle \text{WD}} \approx 0.5295$, whereas for integrable systems obeying Poisson statistics, $\scalebox{1.2}{r}_{\text{Poisson}}\approx0.386$~\cite{Oganesyan_2007,Atas_2013,Corps_2020,Livan_2018,haake2018quantum}. The level spacing ratios calculated for our parameter regimes are shown in Fig.~\ref{fig:levelspacing}. As $L$ increases, all data points approach $\scalebox{1.2}{\text{r}}_{\scriptscriptstyle \text{WD}}$, confirming that the systems reside in the chaotic (thermalized) regime.

\end{document}